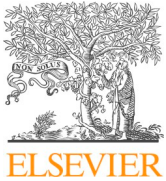

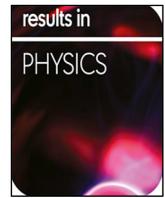

# Charged particle dynamics in the vicinity of black hole from vector-tensor theory of gravity immersed in an external magnetic field

Azeem Nazar[a], Saqib Hussain[b], Adnan Aslam[c,*], Takasar Hussain[d], Muhammad Ozair[d]

[a] Department of Mathematics, School of Mathematics and Computer Science, Hajvery University 33 Industrial Area Gulberg III, Lahore, Pakistan
[b] Institute of Geophysical Astronomy and Atmospheric Sciences (IAG), University of São Paulo (USP), São Paulo, Brazil
[c] School of Electrical Engineering and Computer Science, National University of Sciences and Technology, H-12 Islamabad, Pakistan
[d] Department of Mathematics, COMSATS University Islamabad, Attock Campus, Attock, Pakistan

## ABSTRACT

The dynamics of charged particle in the vicinity of event horizon of a rotating charged black hole (BH) with gravitomagnetic charge from a class of vector-tensor theories of modified gravity immersed in an axially symmetric magnetic field (Bfield) has been studied. The presence of Bfield reduces the radii of innermost stable circular orbit (ISCO) of charged particle and the opposite phenomenon occurs due to the parameter $\beta$ which describes the deviation of modified gravity considered here from the usual Einstein-Maxwell gravity also acts to increase the radius of ISCO. Angular momentum also plays the role regarding the motion of particle which implies that larger the angular momentum easier for a particle to escape. We study the stability of orbits using Lyapunov characteristic exponent which implies more stability in the presence of Bfield. Generally, in the presence of Bfield it is easier for a particle to escape than its absence. We find twist in the trajectories of charged particles close to event horizon due to spin of BH in the presence of Bfield and there is no twist in the absence of it. Finally, we comment on possible utilization of our findings for the relativistic jets and magnetohydrodynamical out flows.

## 1. Introduction

Gravitational wave astronomy gives us new circumstances and ways to study the physics of objects like BH and neutron stars [1]. So that we can investigate the fundamental interplay in a strong gravitational regime and test theories which are alternative to Einstein General Relativity (GR) [2]. The conundrum of dark matter and dark energy impels attempts to modify the GR. Instantly, exploring theories giving accelerating cosmological solution without cosmological constant are really important [3]. There is a very interesting mechanism called screening mechanism [4–6] which can reproduce the results of GR in a weak field, spherically symmetric regime. The study of particle dynamics in the surrounding of BH (solution of new theories of gravity) can provide new test for these theories and also present deviation from GR. There are many interesting results regarding BH solution using screening mechanism and theories (see e. g. [7–9]). Here we focus on vector tensor theory [11,12] and the rotating BH solution obtained in [10]. For a comprehensive review about rotating BH solutions see [13] and for possible deviation of Rotating BH solution (scalar tensor theories) from Kerr family of BHs see [14,15]. It is paramount to study the rotating BH as most of the astrophysical BHs are spinning and theoretically these objects are axially symmetric, (kind of spherical symmetry breaking). Further more, rotating BH can act to accelerate particles from their surrounding up to very high energy. It is concluded that particles can collide in the surrounding of rotating BH and energy grows unbound in their center of mass frame [16–25].

Motion of particles (test particles, charged or neutral) around a BH have a profound significance in the field of astrophysics. This study can lead us to understand the large scale structure of space-time and can explain very energetic phenomenas which occur near the event horizon of BH such as formation of jets (which involve the escaping of particles in the presence of Bfield), accretion disks (particles orbiting around BH), cosmic-rays (CRs) and gamma-rays acceleration. CRs and gamma-rays acceleration are one of the most important area of research in astronomy and the active galactic nuclei (AGN) is one of the prominent candidate to accelerate the ultra high energy CRs [26–28]. The AGN can accelerate the particles with energy $E \sim 10^{20}$ eV [26] and study of these high energy particles can give us some new physical insight about the large scale structure formation, evolution of the active galaxies, cluster of galaxies and universe as well. Blandford-Znajek mechanism explains that the jets can get rotational energy from the BH, which provide the required angular momentum through the accretion of gas and BH mergers [29]. Although the astrophysical evidences for the role of BH spin in gaining energy by jets are still lacking [30]. Accretion disk orbiting a BH generates a general-relativistic field (magnetic-like field) called gravitomagnetic field (G-field) that can accelerates particles upward and inward toward the axis of the disk. The G-field is not the only contrivance to produce the collimated jets. It is one of the G-field effect for a much more tangled situation regarding jet origination. In addition, there is a frame-dragging effect due to the spin of BH which twists the trajectories around the axis hence, playing its part in the formation of the jet.

* Corresponding author.
E-mail address: adnan.aslam@seecs.edu.pk (A. Aslam).

Bfield is ubiquitous in the universe and its intensities may varies in range $(10^{-16} - 10^{15})$G, $(B \sim 10^{-16}$G in cosmic voids and $B \sim 10^{15}$G in Magnetars). Bfield is present in the surrounding of BH which plays the major role to launch a large scale jets [31–33]. Most likely these jets are the source of extremely high energy CRs, coming from extragalactic or galactic region. The origin of the Bfield is probably the existence of plasma in the accretion disk in the vicinity of a BH [34–36]. Bfield only effects the dynamics of charged particle in the surrounding of BH and it does not effect the BH geometry [36]. Static BH with electric charge can produce the electric field which in turn can generate Bfield if it is rotating. This phenomenon can be applied to any kind of stars. Particles in the vicinity of charged rotating BH are under the influence of gravitational and electromagnetic forces which make the problem regarding the particle dynamics more complicated [37–39,51,40].

In this paper, we investigate the dynamics of the particle in the nearby of charged rotating BH immersed in a axially symmetric Bfield. Metric of BH which we are using here is obtained from the "vector-tensor theories of modified gravity valid for arbitrary values of the rotation parameter" [10]. We explore many features of the motion of particle in detail. In first Section 1 we give introduction and importance of the considered problem with some motivation and uses of this research work. In Section 2 introduction of the vector-tensor theory of modified gravity considered in[10] is given. We discuss the basic equations and formalism for the dynamics of the particle in Section 3. Bfield calculations and charged particle formalism is discussed in Section 4. Section 5 is dedicated for ISCO, 6 is for effective potential and 7 is for effective force. We discuss the Lyapunov exponent in Section 8 and escape velocity is discussed in Section 9. In Section 10, we discuss the results and comment on their possible utilizations. Calculations and results are obtained using natural units, $c = 1$, $\hbar = 1$ and $4\pi G = 1$.

## 2. Vector-tensor theory of gravity

We are focusing on the theory called as vector-tensor Galileons [10–12] which have additional vector degrees of freedom associated with dark matter or dark energy model building. These theories have their unique properties and significance for cosmology [61–66], field theory [67–70], and BHs [71–73]. Our aim is to investigate the dynamics of charged particles in the vicinity of BH [10]. The action of standard Einstein-Maxwell system,

$$S_{EM} = \int d^4x \sqrt{-\tilde{g}} \left[ \frac{\tilde{R}}{4} - \frac{1}{4}\tilde{F}^{\mu\nu}\tilde{F}_{\mu\nu} \right].$$ (1)

They [10] have new solutions by using disformal transformation involving vector fields and parameterized by $\beta$,

$$\tilde{g}_{\mu\nu} = g_{\mu\nu}(x) - \beta^2 A_\mu(x) A_\nu(x), \quad \gamma_0^2 = \frac{1}{1 - \beta^2 A^\mu A_\mu}$$

$$\tilde{A}_\mu = A_\mu + \partial_\mu \alpha(x).$$ (2)

For any $\alpha(x)$ the considered transformation has the gauge freedom for the original theory so, after performing the above mentioned modifications the action equation becomes

$$S_{disf} = \int d^4x \sqrt{-\tilde{g}} \frac{1}{4\gamma_0} \left[ R - \frac{\beta^2}{4}\gamma_0^2 \left( S_{\mu\nu}S^{\mu\nu} - S^2 \right) \right.$$
$$\left. - \frac{4 - \beta^2}{4}F_{\mu\nu}F^{\mu\nu} + \frac{\beta^4 - 4\beta^2}{2}\gamma_0^2 F_{\mu\rho}F_\nu^\rho A^\mu A^\nu \right].$$ (3)

Deviates from the usual Einstein-Maxwell case (1) can be recognized by quantities with the disformal parameter $\beta$. This is the action of the modified vector tensor theory, also described in [53,54]. The BH solution given below is obtained by applying the disformal transformation

(2) to Kerr-Newman (KN) solution of action (1) with vector potential profile $A_\mu$(2)[10]

$$ds^2 = \left( \frac{\rho^2}{\Delta\rho^2 - \beta^2Q^2r^2}dr^2 + d\theta^2 \right)\rho^2 - (dt - a\sin^2(\theta)d\phi)^2 \frac{\Delta\rho^2 + \beta^2Q^2r^2}{\rho^4}$$
$$+ ((r^2 + a^2)d\phi - adt)^2\frac{\sin^2(\theta)}{\rho^2},$$
$$\Delta = a^2 + r^2 - 2Mr + Q^2,$$
$$\rho^2 = r^2 + a^2\cos^2\theta.$$ (4)

$$A_\mu = \left( -\frac{Qr}{\rho^2}, \frac{Qr}{\Delta(r)}, 0, \frac{aQr\sin^2\theta}{\rho^2} \right) = \left( A_t, A_r, A_\theta, A_\phi \right)$$ (5)

Resulting geometry obtained after applying the disformal transformation (2) to Kerr-Newman of the Einstein-Maxwell action (1) has naked singularities which are not covered by horizons. But the radial vector profile $A_r(r)$(5) influences the geometry after the disformal transformation. So, the specific profile of black hole is obtained using the radial vector component, which leads to a asymptotically flat black hole configuration (4), [10].

The parameter $\beta$ contributes in the modification of event horizon of BH and play its role in the properties of geodesics of particles. Event horizon and ergo-sphere corresponding to above metric (4) are,

$$g^{rr} = (r^2 + a^2\cos^2\theta)[r^2 - 2Mr + a^2 + Q^2] - \beta^2Q^2r^2 = 0$$ (6)

$$(k^\mu)k_\mu = (r^2 + a^2\cos^2\theta)[r^2 - 2Mr + a^2\cos^2\theta + Q^2] - \beta^2Q^2r^2 = 0.$$ (7)

The above equations have forth order in $r$ so they might have four, two or one real roots. Here we write the external real roots for horizon $r_{Hor}$, and ergo-sphere $r_{ErgoSph}$, respectively,

$$r_{Hor}(\theta = \pi/2) = M + \sqrt{M^2 - a^2 - Q^2(1 - \beta^2)},$$ (8)

$$r_{ErgoSph}(\theta = \pi/2) = M + \sqrt{M^2 - Q^2(1 - \beta^2)}.$$ (9)

So, BH event horizon will change with change of BH charge $Q$ and parameter $\beta$ which differentiate between the geometry of disformal BH from the KN-BH geometry. With the increase of the charge $Q$ for $\beta > 1$ event horizon get smaller. There are possibilities for the naked singularities if $M^2 - a^2 < Q^2(1 - \beta^2)$ but we are not studying them here.

## 3. Neutral particle dynamics

The metric (4) is stationary but non-static since $(dt \to -dt)$ and axially symmetric (invariance under $d\theta \to -d\theta$). The Lagrangian correspond to the BH metric is defined as $(\mathscr{L} = \frac{1}{2}g_{\mu\nu}\dot{x}^\mu\dot{x}^\nu)$ and is independent of $t$ and $\phi$ coordinates which lead to two conserved quantities namely energy $\varepsilon$ and angular momentum $\ell_z$.

Using the Lagrangian formalism [1], we obtained the constant of motion given below,

$$\varepsilon = -g_{tt}\dot{t} - g_{t\phi}\dot{\phi},$$
$$\ell_z = g_{\phi t}\dot{t} + g_{\phi\phi}\dot{\phi}.$$ (10)

Solving Eq. (10), we have,

$$\dot{t} = -\frac{r^2(a^2\varepsilon(r(2m + r) + (\beta^2 - 1)Q^2) - a\ell_z(2mr + \beta^2Q^2 - 1) + \varepsilon r^4)}{a^2(2\beta^2Q^2(-2mr - Q^2 + r^2 + 1) + 4mr(1 - 2mr)}$$
$$+ (Q^2 - r^2)^2 - 1) + r^4(r(r - 2m) + (\beta^2 - 1)Q^2)$$ (11)

---

[1] $\mathscr{L} = 1/2g_{\mu\nu}\dot{x}^\mu\dot{x}^\nu$, the conserved quantities corresponding to $t$ and $\phi$ using Lagrangian equations $\frac{d}{d\tau}\frac{\partial\mathscr{L}}{\partial\dot{t}} = 0$, and $\frac{d}{d\tau}\frac{\partial\mathscr{L}}{\partial\dot{\phi}} = 0$, yielding $\frac{\partial\mathscr{L}}{\partial\dot{t}} = \mathscr{E} \equiv -p_\mu\xi^\mu_{(t)}/m$, and $\frac{\partial\mathscr{L}}{\partial\dot{\phi}} = L_z \equiv p_\mu\xi^\mu_{(\phi)}/m$. One can obtain (11 and 12).

$$\dot{\phi} = \frac{r^2(-a\varepsilon(2mr + \beta^2 Q^2 - 1) + \ell_z r(r - 2m) + (\beta^2 - 1)\ell_z Q^2)}{a^2(2\beta^2 Q^2(-2mr - Q^2 + r^2 + 1) + 4mr(1 - 2mr)}$$
$$+ (Q^2 - r^2)^2 - 1) + r^4(r(r - 2m) + (\beta^2 - 1)Q^2) \tag{12}$$

From the normalization condition, $u^\mu u_\mu = -1$ with

$$u^\mu = \left(\frac{dt}{d\tau}, \frac{dr}{d\tau}, \frac{d\theta}{d\tau}, \frac{d\phi}{d\tau}\right), \tag{13}$$

we get the equation for $\dot{r}$,

$$\frac{r^3}{a^2 + r(r - 2M) + (1 - \beta^2)Q^2}\left(\frac{dr}{d\tau}\right)^2 + r^2 + \left(\ell_z - a\varepsilon\right)$$
$$- \frac{((a^2 + r^2)\varepsilon - a\ell_z)^2}{r(a^2 + r(r - 2M) + (1 - \beta^2)Q^2)} = 0. \tag{14}$$

Here we are doing the calculation for equatorial plane ($\theta = \frac{\pi}{2}) \Rightarrow \dot{\theta} = 0$ and we can rearrange the above equation as

$$\frac{1}{2}\left(\frac{dr}{d\tau}\right)^2 + \frac{1 - \varepsilon^2}{2} = V_{eff}\left(r, \varepsilon, \ell_z\right) \tag{15}$$

and $V_{eff}$ is the effective potential for the considered geometry of BH,

$$V_{eff}\left(r, \varepsilon, \ell_z\right) = \frac{a^2(1 - \varepsilon^2) + \ell_z^2 + (1 - \beta^2)Q^2}{2r^2} + \frac{(1 - \beta^2)(\ell_z - a\varepsilon)^2 Q^2}{2r^4} - \frac{M}{r} - \frac{M(\ell_z - a\varepsilon)^2}{r^3} \tag{16}$$

The Eq. (15) can be solved analytically for a short distance approach (neglecting terms with $1/r$ and $1/r^2$ dependence) or for long distance approach (neglecting terms involving $1/r^3$ and $1/r^4$) but not in general. The solution of the Eq. (15) is fully dependent on the $V_{eff}$ effective potential (16). So, we are analyzing Eq. (16) for co-rotating ($\ell_z > 0$) and counter-rotating ($\ell_z < 0$) trajectories of particle with respect to BH. Effective potential expression (16) has a identical structure to the KN case for $\beta \to 0$[74]. For $\beta^2 > 1$, the signs among different terms to the effective potential (16) is not same as compared to standard KN-BH [75,76]. So, this different regime may lead to qualitatively new features for circular orbits which will be discuss in Section (6).

There are three possibilities for a particle having collision while moving in a circular orbit in the surrounding of BH: (i) remain bounded in the orbit with little perturbation, (ii) fall into the BH, (iii) escape to infinity. Suppose there is a small perturbation in the motion of particle after collision and it attains new values of energy $\varepsilon$, azimuthal angular momentum $\ell_z$, and the total angular momentum defined as.

$$\ell^2 = r^4(\dot{\theta}^2 + \sin^2\theta \; \dot{\phi}^2) \qquad = r^2 v_\perp^2 + r^4 \sin^2\theta \dot{\phi}^2, \quad v_\perp = r\dot{\theta}. \tag{17}$$

We incorporate these changes with, azimuthal angular momentum $\ell_z$ goes to total angular momentum $\ell$ and effective potential becomes the function of total angular $V_{eff}(r, \varepsilon, \ell_z) \to V_{eff}(r, \varepsilon, \ell)$. After collision the particle may gain a velocity ($v_\perp$), orthogonal to the equatorial plane [52].

## 4. Magnetic field and charged particle

### 4.1. Magnetic field

It is of extreme importance to investigate the electromagnetic fields and particle dynamics in vector-tensor theories with the aim to get new tool and physical insights for studying new relativistic effects. This kind of demonstration and research can be interesting because of the existence of evidences that a Bfield exists in the surrounding of BHs. Here, we use the weak Bfield limit in the sense that the energy and momentum of this field can not change the background geometry of BH [42]. According to this condition for a BH with mass $M$, the strength of Bfield should satisfy the following relation,

$$B \ll B_{max} \sim 10^{19}\frac{M_\odot}{M_{BH}} Gauss. \tag{18}$$

This condition satisfied by the stellar BH and supermassive BH as well. The BH which satisfies the condition (18) is known as "weakly magnetized" ($\Rightarrow B \sim (10^4 - 10^8) Gauss \ll 10^{19}$) [43]. The Bfield may provide energy to the particles moving near to event horizon of the BH for their escape to infinity [44–47,51].

The considered spacetime geometry (disformal-BH) has Killing vectors $\xi^\mu = (\partial/\partial t)^\mu$ and $\psi^\mu = (\partial/\partial\varphi)^\mu$. Therefore, vector potential can be written as a linear combination of these Killing vectors. For a neutral rotating BH the vector potential is defined as [[56]],

$$A^\mu = \frac{\mathscr{B}}{2}\left(\psi^\mu + 2a\xi^\mu\right) \tag{19}$$

This type of $A^\mu$ yields an asymptotically uniform Bfield with strength $\mathscr{B}$. For a rotating BH with charge $Q$ Eq. (20) becomes [32,55]

$$A^\mu = \frac{\mathscr{B}}{2}\left(\psi^\mu + 2a\xi^\mu\right) - \frac{Q}{2M}\xi^\mu \tag{20}$$

The term with $a$ corresponds to the effect of Faraday induction due to the BH rotation which leads to accretion of positively charged particles towards BH horizon [57]. Accretion remains until the potential difference vanish and the BH will gain a charge of $Q = 2aM\mathscr{B}$[58]. As the accretion vanishes then the vector potential becomes

$$A^\mu = \frac{\mathscr{B}}{2}\psi^\mu \tag{21}$$

To satisfy the condition (18), the induced charge of the BH should be like $Q/M \leqslant 2\mathscr{B}M \ll 1$ so that its effect on BH geometry can be neglected. Therefore, we choose the 4-vector potential given in Eq. (21). The vector potential is invariant under the symmetries of considered BH metric, i.e.,

$$L_\xi A_\mu = A_{\mu,\nu} \; \xi^\nu + A_\nu \; \xi^\nu_{,\mu} \tag{22}$$

The Bfield is defined as [59]

$$\mathscr{B}^\mu = -\frac{1}{2}e^{\mu\nu\lambda\sigma}F_{\lambda\sigma}u_\nu, \tag{23}$$

with

$$e^{\mu\nu\lambda\sigma} = \frac{\epsilon^{\mu\nu\lambda\sigma}}{\sqrt{-g}}. \tag{24}$$

where (24)$\epsilon^{\mu\nu\lambda\sigma}$ is the Levi Civita symbol and the Maxwell tensor is defined as

$$F_{\mu\nu} = A_{\nu,\mu} - A_{\mu,\nu}. \tag{25}$$

The Bfield configuration is similar to [51] with the change of integral of motion with respect to the considered geometry.

### 4.2. Charged particle dynamics

The Lagrangian of particle moving in the Bfield in the vicinity of BH is given by (see [32] and references given therein)

$$\mathscr{L} = \frac{1}{2}g_{\mu\nu}\dot{x}^\mu\dot{x}^\nu + \frac{qA_\mu}{m}\dot{x}^\mu, \tag{26}$$

and generalized 4-momentum of the particle is, $P_\mu = mu_\mu + qA_\mu$. Here $q$ and $m$ are the charge and mass of particle. Now the conserved quantities (constant of motion) for charged particle will remain conserved with little modification due to the Bfield,

$$\varepsilon_B = \varepsilon + 2ab = -g_{tt}\dot{t} - g_{t\phi}\dot{\phi}$$
$$\ell_b = \ell_z - b = g_{\phi t}\dot{t} + g_{\phi\phi}\dot{\phi}, \quad b = \frac{q\mathscr{B}}{2m}. \tag{27}$$

Solving the Eq. (27) we have

$$\dot{t} = -\frac{(2ab+\varepsilon)(a^2(r(2m+r)+(\beta^2-1)Q^2)+r^4)+a(b-\ell_z)}{(2mr+\beta^2Q^2-1)} \Bigg/ r^2\left(\frac{a^2(2\beta^2Q^2(-2mr-Q^2+r^2+1)+4mr(1-2mr)+(Q^2-r^2)^2-1)}{r^4}\right.$$
$$\left.+ r\left(r-2m\right)+(\beta^2-1)Q^2\right)$$

(28)

$$\dot{\phi} = \frac{r^2(-2a^2b(2mr+\beta^2Q^2-1)-a\varepsilon(2mr+\beta^2Q^2-1)-(b-\ell_z)}{a^2(2\beta^2Q^2(-2mr-Q^2+r^2+1)+4mr(1-2mr)} \cdot \frac{(r(r-2m)+(\beta^2-1)Q^2))}{+(Q^2-r^2)^2-1)+r^4(r(r-2m)+(\beta^2-1)Q^2)}$$

(29)

The constants of motion regarding the charged particle in the presence of Bfield are modified (particle energy $\varepsilon_B$(28) and angular momentum $\ell_B$(29)). Using the normalization condition $u^\mu u_\mu = -1$ and Eqs. (28) and (29), we can obtain the equation for $\dot{r}$ and effective potential $V_{eff}$. So, the radial equation and the effective potential with the consideration of Bfield become

$$\frac{1}{2}\left(\frac{dr}{d\tau}\right)^2+\frac{1-(\varepsilon+2ab)^2}{2}=V_{eff}\left(r,b,\ell,\varepsilon,a,Q,\beta\right)$$

(30)

$$V_{eff}\left(r,b,\ell,\varepsilon,a,Q,\beta\right)=\frac{a^2(1-(2ab+\varepsilon)^2)+(\ell^2-b)^2+(1-\beta^2)Q^2}{2r^2}$$
$$-\frac{M(-a(2ab+\varepsilon)-b+\ell)^2}{r^3}$$
$$+\frac{(1-\beta^2)Q^2(-a(2ab+\varepsilon)-b+\ell)^2}{2r^4}-\frac{M}{r}.$$

(31)

Here, we consider the small change in angular momentum and energy of the charged particle (same analogy as we consider for neutral particle) when it collides with another particle and the total angular momentum is given in the Eq. (17).

So, the Bfield presence leads the charged particle under the effective potential given by Eq. (33) which is discussed in detail in the next section. Note that as the Bfield does not effect the geometry, therefore, the major effect on the dynamics of the charged particle is comes from the G-field (curve geometry of disformal BH) specially near the event horizon and ISCO.

## 5. Innermost stable circular orbit (ISCO)

Here, we are examining the properties of marginally stable circular orbits of particle. Such trajectories are called the ISCO [60,51]. The whole analysis is done using the dimensionless parameters which are given as

$$e=\frac{\varepsilon}{m},\quad b=\frac{q\mathscr{B}}{2m},\quad L=\frac{\ell}{m}$$

(32)

and the BH dimensionless parameters are, $a=\frac{J}{M_{BH}^2}$ and $Q=\frac{Q}{M_{BH}}$. For the circular orbit of specific radius, both the radial velocity ($\frac{dr}{d\tau}=0$) and the acceleration $\left(\frac{d^2r}{d\tau^2}=0\right)$ should vanish. So the effective potential in the absence of Bfield after using the dimensionless parameters become,

$$V_{eff}\left(r,L,e,a,Q,\beta\right)=\frac{a^2(1-e^2)+L^2+(1-\beta^2)Q^2}{2r^2}+\frac{(1-\beta^2)(L-ae)^2Q^2}{2r^4}$$
$$-\frac{M}{r}-\frac{M(L-ae)^2}{r^3}\equiv\frac{1-e^2}{2},$$

(33)

in the presence of Bfield it is,

$$V_{eff}\left(r,b,L,e,a,Q,\beta\right)=\frac{a^2(1-(2ab+e)^2)+(L-b)^2+(1-\beta^2)Q^2}{2r^2}$$
$$-\frac{M(-a(2ab+e)-b+L)^2}{r^3}+\frac{(1-\beta^2)Q^2(-a(2ab+e)-b+\ell)^2}{2r^4}$$
$$-\frac{M}{r}=\frac{1-(e+2ab)^2}{2}.$$

(34)

To get the stability condition for the orbits we will use the first and second derivative test of effective potential,

$$\left(\frac{\partial V_{eff}(r,L,e,a,Q,\beta)}{\partial r}\right)\Bigg|_{r=r_0}=0,\quad\left(\frac{\partial^2 V_{eff}(r,L,e,a,Q,\beta)}{\partial r^2}\right)\Bigg|_{r=r_0}\geqslant 0$$

(35)

We have solved the above Eqs. (33), (35) for $e^2$ and $L^2$ in the absence of Bfield,

$$e_o^2=\frac{q(3-4r_0)+r_0(3r_0-2)}{r_0(3r_0-4q)},$$
$$L_o^2=\frac{a(3q-2r_0)+r_0(4q^2-9qr_0+6r_0^2)}{r_0(3r_0-4q)},$$

(36)

with $q=Q^2(1-\beta^2)$. In the presence of Bfield constant of motion and effective potential have changed as given in Eqs. (27) and (34). Hence, the energy and angular momentum correspond to the ISCO also modified accordingly. Solving Eqs. (33) and (35) we find the critical expressions for the Bfield $b_o$ and angular momentum $L_o$ correspond to ISCO as:

$$L_o=\frac{1}{r_0(3-r_0)+2q}\left[b_o r_0((6a^2+3)-r_0)+2(2a^2+1)qb_o+3ae_o r_0+2aqe_o\right.$$
$$\pm r_0(4a^4b_o^2r_0^2+4a^3b_o e_o r_0^2+a^2(r_0((e_o^2-1)r_0+3)+2q)$$
$$\left.-(r_0+q)(r_0(3-r_0)+2q))^{\frac{1}{2}}\right],$$

(37)

$$b_o=\frac{1}{(2a^2+1)(6a^2r_0(4+r_0)+10(2a^2+1)q+12r_0-3r_0^2)}$$
$$[-2ae_o(6r_0+5q)+3L_o r(4-r_0)+10qL_o$$
$$+4a^2L_o(6r_0+5q)-2a^3e_o(3r_0(4+r_0)+10q)\pm r_0(36a^3e_o L_o r_0^2$$
$$-12a^5(3r_0(4+r_0)+10q)+(2r_0+3q)(3r_0(4-r_0)+10q)$$
$$+a^2(3r_0(3(e_o^2-1)r_0+4(3-8r_0))-2q(112r_0-15)-120q^2)$$
$$+4a^4(9L_o^2r_0^2-q(56r_0+9r_0^2-30)-6r_0(4r_0+r_0^2-6)-30q^2))^{\frac{1}{2}}].$$

(38)

Using the values of $L_o$ and $b_o$ in Eq. (34) we have $e_o$.

For radial position $r_o$ of the ISCO we have an algebraic equation of sixth degree [10]. There can be six real roots for different value of parameters like $a,Q,\beta$. But, the ISCO should be outside the event horizon. We can solve first derivative of the effective potential (34) to get the critical vlues for ISCO radius. But that equation is the fifth order in $r$ with six parameters which can not be solved analytically. So, we solve it numerically. From these numerical solution we plot the ($r_o/r_h$) for different parameters as shown in Fig. 1. It is shown by Fig. 1 that $r_o$ decreases with the increase in Bfield strength indicating that the Bfield presence acts to decrease the radius of $r_o$ but it increases with increase of disformal parameters $\beta$. Fig. 1 also shows that the $r_o$ is greater for retrograde orbits ($L=-5$) than prograde orbits ($L=5$). For the BH charge, $r_o$ gets smaller for larger value of $Q$.

## 6. Effective potential

Here we are discussing the different aspects of the dynamics of charged and neutral particles. Effective potential explains the behavior of particles in the vicinity of BH where they are influenced by the BH gravity and electromagnetic force. As we are considering the vector-tensor theory of gravity so we will discuss the difference between the general relativity and the considered one. Effective potential Eq. (33) has four terms with different $r$ dependence as ($r^{-1}$, $r^{-2}$, $r^{-3}$, $r^{-4}$). These terms depend on the values of parameters which can make them

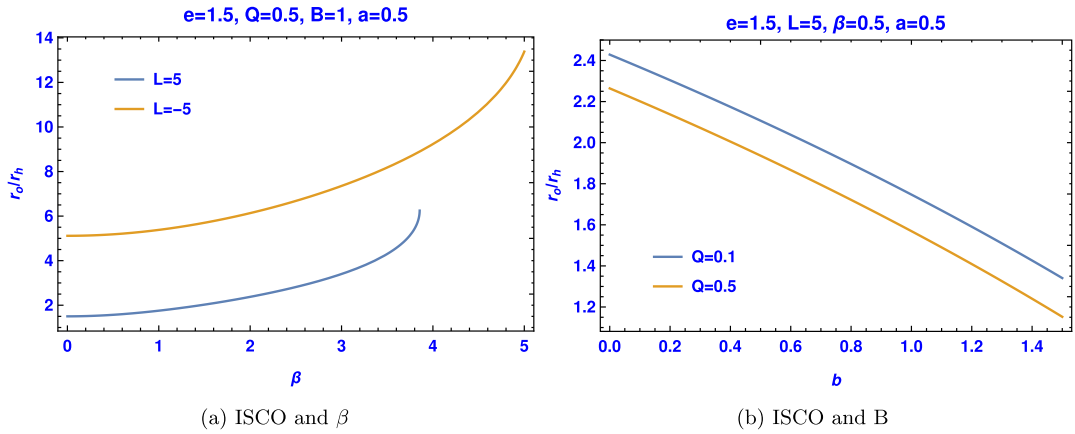

**Fig. 1.** Panel (a): represents the change in ISCO radius ($r_o/r_h$) with respect to disformal parameter $\beta$ for different value of angular momentum. Panel (b): show the $r_o/r_h$ vs Bfield for different value of BH charge $Q$.

attractive or repulsive. We are studying the effective potential for different values of parameters to explore their consequences on the dynamics of particles moving in the BH vicinity. We plot $V_{eff}(r, \epsilon, L)$ in Fig. 2 and 3 for different values of angular momentum (L), and explain the conditions on energy of the particle needed to escape or remain bound in an orbit, around the disformal-BH. For a particle moving in the surrounding of disformal-BH under the effective potential shown in the Fig. 2 has three possibilities:

[i] Bound motion that particle moving in an orbit around the BH,
[ii] Captured or fall into the BH,
[iii] Escape to infinity.

For a particle coming from infinity having large L, needs more energy to climb the effective potential and then fall into the BH as compared to the particle with less L. It can be seen from Fig. 2 that for $L = 6$, particle need more energy to climb the effective potential and fall into the BH than $L = 5$. So, the possibility for a particle with less L to fall into the BH is greater as the potential barrier is smaller for the case of large L and vice versa. Depending on the energy of the particle it can have bound motion that it can move in a stable orbits around the BH.

There are two kind of motion with respect to L, negative ($- L$) is for corotating and positive ($+ L$) is for counter rotating orbit with respect to the BH spin. Here, we will study the behavior of orbits for different parameters corresponding to the co-rotating and counter-rotating disformal BH. In general, particle does not have a bound motion it may rotate around the BH for a while then escape or capture after that due to different phenomenon going on in the vicinity of BH. So, we also discuss the time period for a particle to remain in the bound orbits.

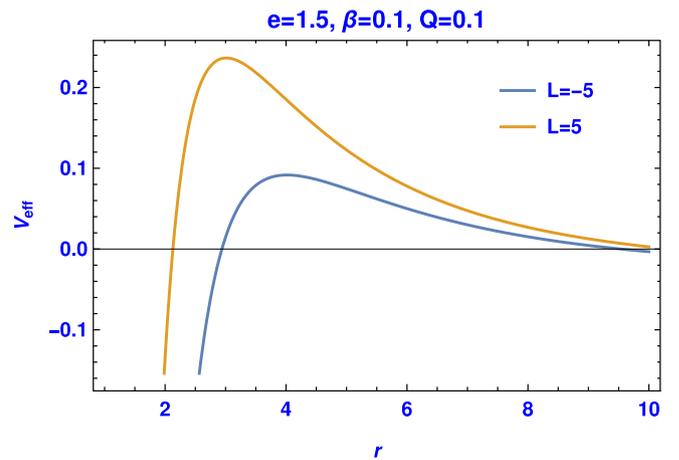

**Fig. 3.** Effective potential vs r for different value of angular momentum L, retrograde and prograde orbits.

Figs. 4 and 5 show the behavior of effective potential for different values of spin parameter of the BH. From Fig. 4, it can be seen that co-rotating particle can come more closer to the horizon as compared to counter-rotating, in agreement with [10]. In Fig. 5 we study the general response of effective potential for BH spin $a$. In general, as the BH spin increases, the ISCO exists away from the event horizon due to frame dragging effect with the spin which is presented in Fig. 5. It shows that location of ISCO is far from the horizon for a BH with high spin rate. From Fig. 5 one can analyze that possibility for the ISCO to exist near horizon is low. It is a sort of shift in the ISCO position with the increase of spin.

Event horizon and ergo-sphere for the disformal BH are larger then the Kerr-Newman BH as shown in Figure-1 of [10]. With the increase of parameter $\beta$, the force of attraction also increases for a particle moving in the vicinity of BH as shown in Fig. 6. So, the capturing possibilities for the particle increase and the position of ISCO shifts away from the BH event horizon. In Fig. 6 for $\beta^2 = 3$ we use $Q = 0.1$ and $\beta^2 = 6$ correspond to $Q = 1$ (extreme condition) so, for both cases ISCO is far from the event horizon. But for $\beta = 0$ the ISCO exists very near to event horizon.

In Fig. 7 we are exploring the Bfield contribution toward effective potential conduct. There is a clear change of behavior of effective potential for different value of Bfield strength as shown in Fig. 7. For $b = 2$ & $b = 1$, there is a clear shift of minimum value $V_{min}$ (local minimum value of $V_{eff}$) of effective potential toward BH horizon which (8). In the existence of charge on the BH the minimum of effective potential (which corresponds to the ISCO) shifts away from the BH indicating that the orbits of charged particles may become unstable. So, the radius of stable orbits increases due to charge of BH. The influence of Bfield has the opposite effect. The Bfield acts to decrease the radius of the orbits which means that the charged particles may come much closer to the BH. So, the radius of ISCO decreases due to the presence of Bfield and increases due to the charge of BH.

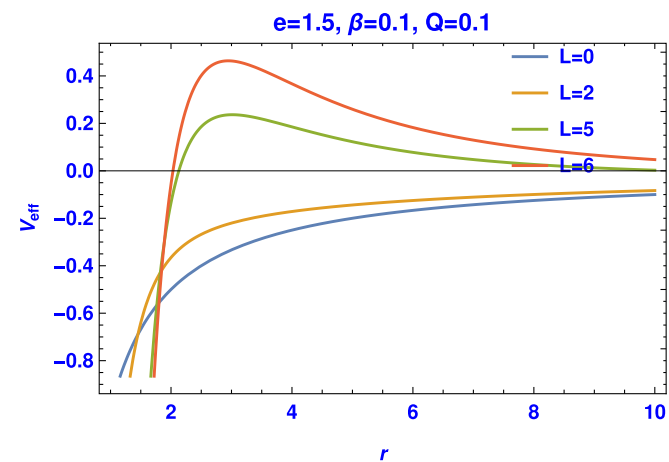

**Fig. 2.** Effective potential vs r for different values of angular momentum L.

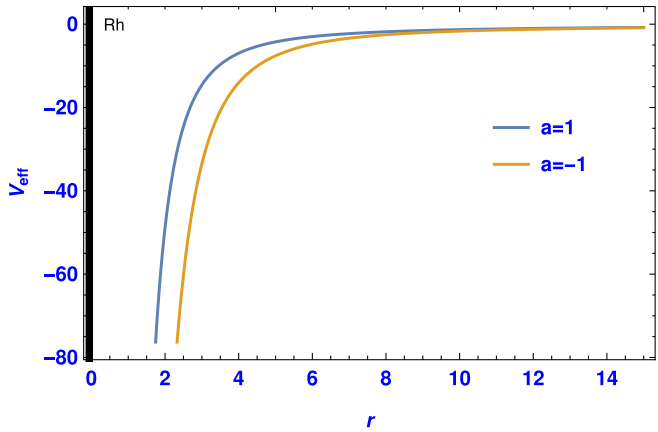

**Fig. 4.** This figure shows the behavior of effective potential with respect to prograde and retrograde orbits of the particle moving around disformal BH. So, for $a = 1$ (prograde) ISCO of the particle can be more closer to BH-event horizon than retrograde, $a = -1$.

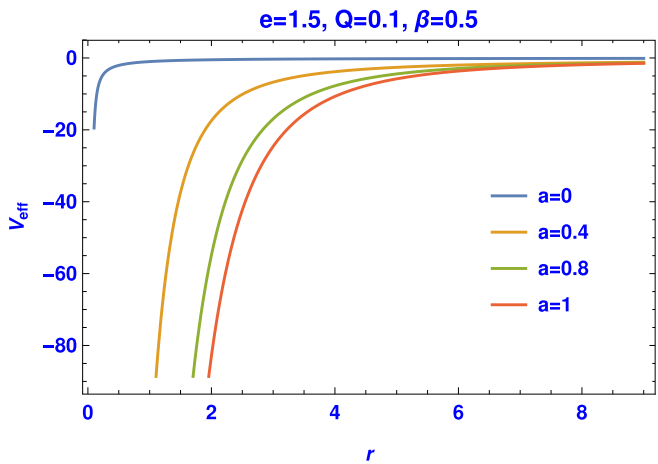

**Fig. 5.** Effective potential vs r for different value of the BH spin.

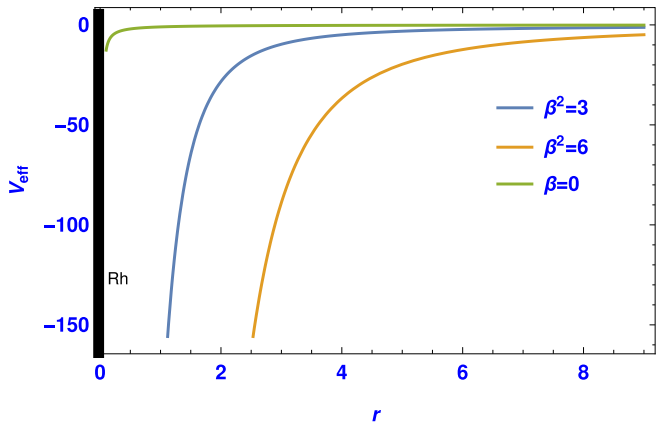

**Fig. 6.** Effective potential vs r for different value of parameter $\beta$.

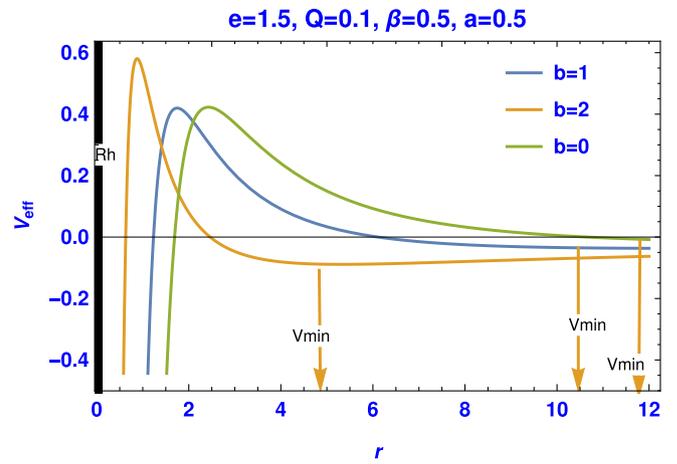

**Fig. 7.** Effective potential vs r for different value of Bfield.

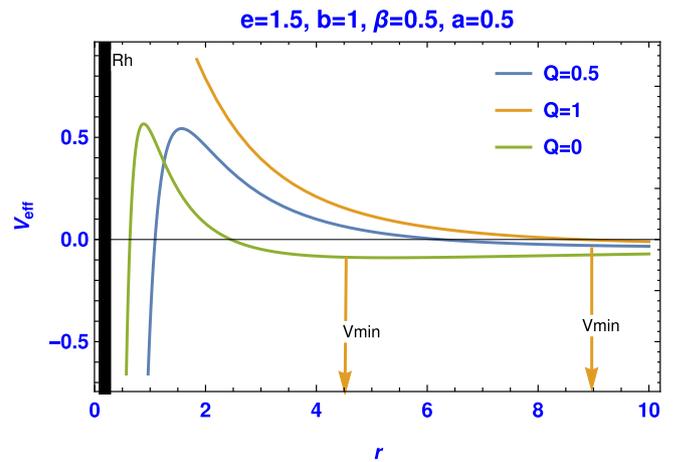

**Fig. 8.** This figure shows the shift of ISCO away from the horizon, $V_{min}$ correspond to ISCO. Here we plot BH potential for different values of BH charge $Q$.

## 7. Effective force on the particle

The effective force on the particle moving in the surrounding of BH is defined,

$$F = \frac{-1}{2} \nabla_r \left( V_{eff}\left( r, \varepsilon, \ell_z \right) \right), \quad \nabla_r = \frac{\partial}{\partial r}. \tag{39}$$

Similarly, we can write for charged particle,

$$F = \frac{-1}{2} \nabla_r \left( V_{eff}\left( r, \varepsilon_b, \ell_b \right) \right). \tag{40}$$

clearly indicate the repositioning of the ISCO toward horizon. Generally, the ISCO radii of charged particles orbiting around BH decreases in the presence of the axially symmetric Bfield. The opposite phenomenon occurs when the BH has a charge. The neutral particle coming from infinity collides with charged particle moving in the ISCO, the collision energy is decreased by the charge of the BH in the presence of Bfield.

BH-Event horizon get shrink with the growth of charge $Q$, see Eq.

$$
\begin{aligned}
F &= \frac{1}{2}\Bigg[ -\frac{3M(-a(2aB+\varepsilon)-B+\ell_z)^2}{r^4} \\
&\quad + \frac{2(1-\beta^2)Q^2(-a(2aB+\varepsilon)-B+\ell_z)^2}{r^5} - \frac{M}{r^2} \\
&\quad + \frac{a^2(1-(2aB+\varepsilon)^2)+(\ell_z-B)^2+(1-\beta^2)Q^2}{r^3} \Bigg]
\end{aligned}
\tag{41}
$$

After the collision, $\ell_z \to \ell$ and the force $F(r, \varepsilon_b, \ell_z) \to F(r, \varepsilon_b, \ell)$ which is the force acting on the charged particle in the vicinity of disformal-BH (4) in the presence of Bfield. This force includes all the curvature effects due to BH mass (M), spin (a) and it also incorporate the Bfield contribution. In the presence of Bfield, the force on the particle gets more attractive and particle can be more closer to event horizon, so, radius of the ISCO decrease, see Fig. 9. But at the same time, particle may get more energy due to the Bfield which may leads it to easy escape so, capturing possibilities for a particle by the BH decrease. For $b = 1$ implies that the possibilities for particles to escape from BH surrounding increase and ISCO radius decrease in the presence of Bfield as shown in Fig. 9. For very large distance from the BH, it can be approximated with Lorentz force (charged particle in a Bfield ($F = q(v \times B)$)). It is quite important to compare the strength of the gravitomagnetic (BH-spin, BH-curvature, angular momentum of accretion disk) effect to that of magnetohydrodynmics (MHD) driven outflows. Acceleration via an MHD electromagnetic-processes such as the Blandford and Znajek effect is dependent on the strength of the Bfield and the angular momentum $L$ of the accretion disk. Acceleration due to Bfield continues at distances much farther from the BH, while the gravitomagnetic effect can only be important near the ISCO and accretion disk during the initial stages to launch the jet.

## 8. Lyapunov exponent for the instability of Trajectories

In mathematics the Lyapunov characteristic exponent (LE) of a dynamical system is a quantity which identifies the rate of separation of infinitesimally close trajectories. It can be also used to study the behavior of the system either it is chaotic or stable. So, it can be used to investigate the stability of circular orbits. A positive LE is usually taken as an indication that the system is chaotic. Fig. 10 shows that the value of LE is positive near the horizon which corresponds to highly chaotic behavior of the system and system is comparatively less chaotic in the Bfield presence than its absence,

$$
\sqrt{\frac{-V''_{eff}(r, \varepsilon, \ell)}{2i(r, \varepsilon, \ell)}}, \quad V''_{eff}\left(r, \varepsilon, \ell\right) = \frac{\partial^2 V_{eff}(r, \varepsilon, \ell)}{\partial r^2}.
\tag{42}
$$

For $b = 1$ and $b = 0$, LE is zero away from the horizon (for large distance) which correspond to stability of orbits. It can be seen from Fig. 10, near the horizon value of LE goes to zero quickly in the presence of Bfield $b = 1$ than its absence $b = 0$ which means that the Bfield reduce the ISCO radius.

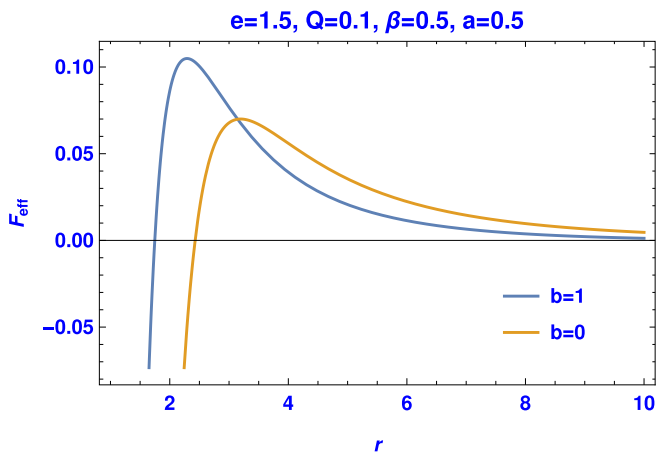

**Fig. 9.** Effective force vs r for different value of Bfield b.

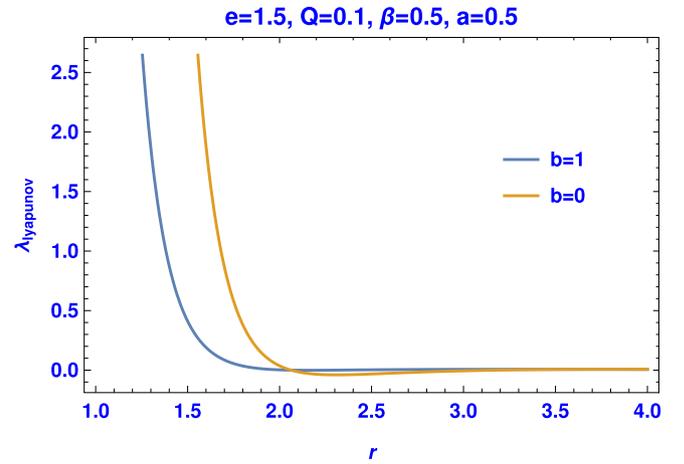

**Fig. 10.** Lyapunov exponent vs r for different value of Bfield b.

## 9. Escape velocity and trajectories of particle

We have calculated the escape velocity of a particle orbiting the disformal-BH and immersed in a Bfield (weakly magnetized-BH (Frolov-2012)) [32]. Comparison is also done for escape velocity against different parameters (Bfield (b) and charge of BH(Q)) Fig. 11. Charged particle moving in a magnetized environment may face a magnetic shock by which it can gain enough energy to escape from that environment. There are many models which explain the shock acceleration and propagation of charged particle (e.g. CRs) in the presence of magnetic field [77,78] in different turbulent magnetized media. This topic is beyond the scope of this article, so we just mentioned it here. Hence the high b strength corresponds to easy escape of particle from BH surrounding. Escape velocity is larger for large b therefore the possibility for a particle to escape from the BH surrounding increases with the increase of b as shown in Fig. 11. Horizon does not change with the change in b strength, panel (a) of Fig. 11. In case of $Q$, BH horizon shrink as $Q$ increases so, particles can escape easily as shown in panel (b) of Fig. 11. Escaping of particle from the BH vicinity is one of the phenomenon which can contribute in the formation of jets.

The changes in the trajectories of particles, that occur because of the curved geometry, can have direct influence on the electromagnetic and plasma processes near bodies like BH [79]. There are many factors which contribute in the origination of collimated jets including the Blandford and Zanek effect [29]. As most of the BHs have spin [49] so, one should consider the Gravitomagnetic effects of the spinning BH and the frame dragging effect due to the angular momentum of the accretion disk [48]. Gravitomagnetic acceleration with the BH spin may twist the trajectories near the ISCO or horizon. From Fig. 12 (panel (a) and (b)) one can see the twisting of trajectories which indicate the role of spin of the BH.

Geodesic equations for particles are

$$
\frac{dt}{dr} = \frac{dt}{d\tau} \times \frac{d\tau}{dr}
\tag{43}
$$

$$
\frac{d\phi}{dr} = \frac{d\phi}{d\tau} \times \frac{d\tau}{dr}
\tag{44}
$$

Geodesics of particles, temporal $t(r)$ and angular $\phi(r)$ are the solution of Eqs. (43) and (44). We solve these equation numerically using **NDSolve** command of **Mathematica**. So, trajectories of the particles are shown in the Fig. 12. We also compare geodesics of particles for different kind of geometry (e.g. Schwarzschild-BH, disformal-BH) in panel (b) of figure Fig. 12. The twist in the trajectories of the particles, for $b = 1$ occurs much closer to horizon than for $b = 0$ which also indicate that particles can come close to BH horizon in the presence of Bfield as shown in panel (a) of Fig. 12. Panel(b) of Fig. 12 shows that there is no twist in the trajectories for the S-BH even in the presence of b. Panel (c) show the change in the angular trajectories due to (b). The Bfield may give enough energy for a particle to escape from the BH-potential. Event horizon change with the change in spin and charge of the BH, so as the geodesics of the particles. Particles can be compelled more closer to the BH due to the Bfield. As the disformal BH has charge and it is also spinning so, trajectories of particles get twisted.

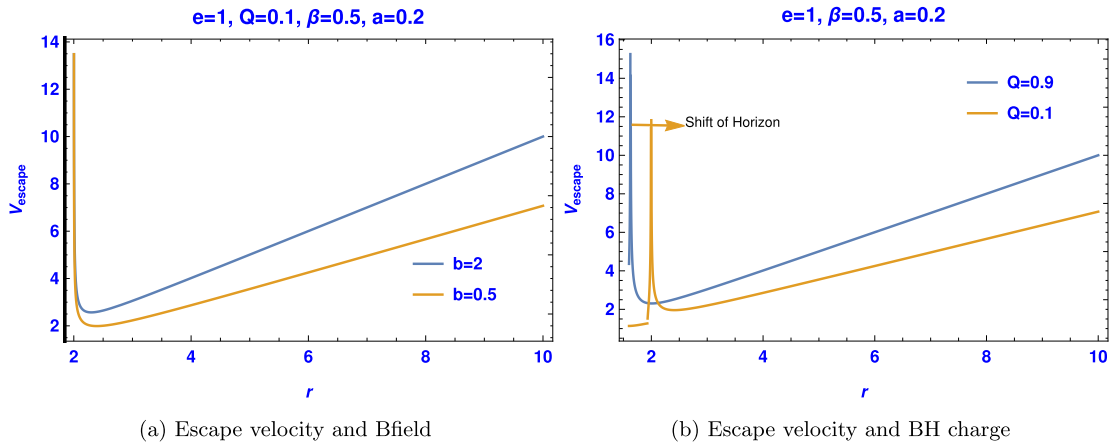

**Fig. 11.** Escape velocity of the particle for different value of BH charge ($Q$) and Bfield ($b$).

Time for a particle to stay in an orbit around the BH is studied in Fig. 13. In general the motion of the particle around the BH is unstable, so, time to stay in an orbit is finite. Time ($T_{orbit}$) for a disformal BH (like KN-BH) is much less than a BH like the Schwarzschild-BH. It can be seen from Fig. 13, for $a = 0$, time period for a particle to be in an orbits is increasing with distance but for $a = 0.5$ and $a = 1$, it attain the maximum value and then goes to zero. for the extreme case $a = 1$ time for a particle to remain in an orbit is large than the case of $a = 0.5$. Larger the spin larger will be the time but always less from the static BH like Schwarzschild BH. Although the particles time to remain in an orbits around the BH is always finite, either they escape from the BH vicinity or captured by the BH.

## 10. Discussion and conclusions

Over the past decades, many detailed studies have been carried out to explore the structure and properties of electro-magnetic fields of astrophysical objects, their interaction with different turbulent media they are transiting through and their behavior in the presence of gravitational fields.

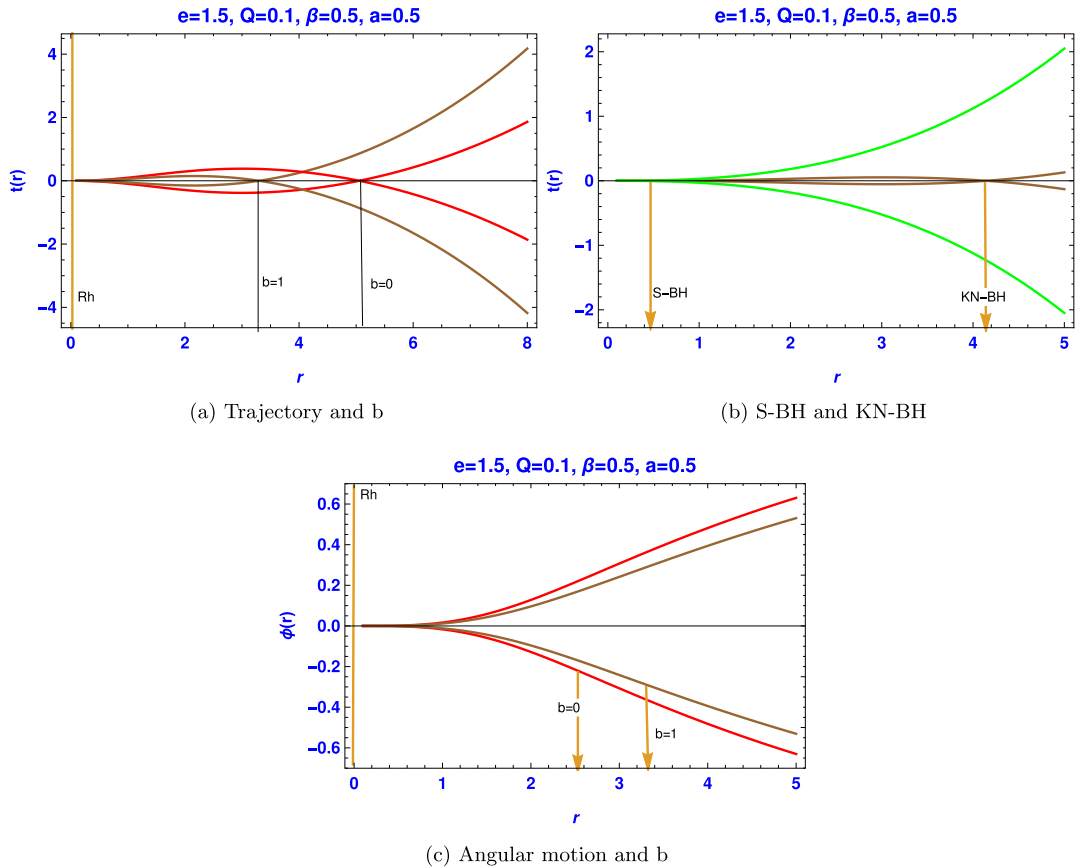

**Fig. 12.** This FIG. shows the comparison of the trajectories of particles for different kind of geometry (e.g. Schwarzschild-BH, disformal-BH) in the presence of Bfield.

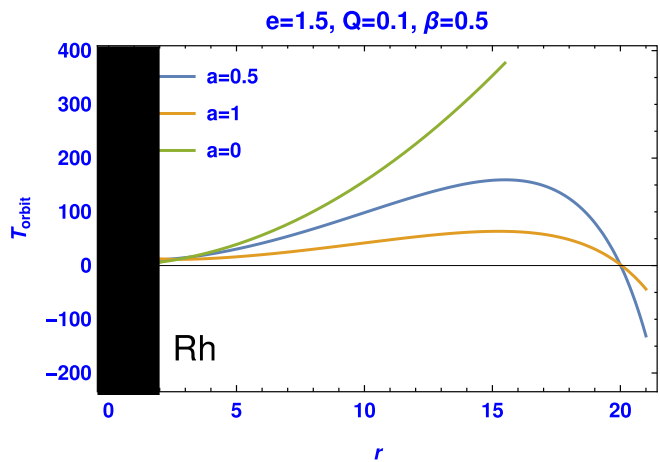

**Fig. 13.** Time period of a particle in circular orbits around the BH, for different values of BH spin $a$ parameter.

Here, we compared the gravitational and the electromagnetic fields close to compact objects and their impact on the motion of charged particles. Dynamics of the particles (neutral and charged) have been studied in the vicinity of weakly magnetized disformal-BH and we have also compared different features of motion with other BH-solutions (S-BH, Kerr-BH, KN-BH). We have tried to address the phenomenon of formation of relativistic jets, although it is a very complicated problem but the different features of particle motion we studied here are very relevant to the jets. Specially the processes, the Bfield contribution to accelerate the charged particle, twisting of the trajectories due to spin of BH and Bfield effect on the radius of ISCO is very important to understand their formation.

With the increase of parameter $\beta$, the force on the particle becomes more attractive which enhances the capturing possibilities for a particle. It is found that the ISCO position shifts away from BH-horizon with the increase of disformal parameter $\beta$ and the opposite phenomenon occurs due to the presence of the Bfield, shown in Figs. 1 and 6). From our analysis we established that in the presence of Bfield, particle get more closer to event horizon result in reduction of the radius of ISCO. But at the same instance particle get more energy from the Bfield which leads to easy escape see Figs. 7 and 9. We found that the Bfield and disformal parameter are acting opposite regarding the position of ISCO. It is also concluded that the ISCO radius get reduced as the BH-charge increase, Fig. 8.

We established that the particle motion is largely depend on the angular momentum that particle with large L can escape easily from the BH vicinity. It is also found that the ISCO radius is large for the case retrograde motion than prograde see Figs. 1, 2 and 3. Greater the spin larger will be the time of the particle to stay in an orbit around the disformal-BH (like KN-BH) but always less than the static BH like Schwarzschild-BH, Fig. 13. Lyapunov exponent indicated that the orbits near the event horizon are highly chaotic and more stable in the existence of Bfield $b = 1$ than its nonexistence $b = 0$.

Escape velocity of the orbiting particle is calculated which indicates that the escape possibility of a particle from the BH surrounding increases with the increase of Bfield. It is also presented that the event horizon shrinks with the increase of BH charge, Fig. 11.

The changes in the trajectories of charged particles which occurred because of the curved geometry specially, due to the spin of the BH. We found the twist in the trajectories of the particles occur much closer to event-horizon in the existence of Bfield than its absence which also indicates that particles can come close to the BH horizon in the presence of Bfield. Comparison of the trajectories is done for S-BH (BH like-Schwarzschild) and KN-BH (disformal-BH or BH like Kerr-Newmann), there is no twist for the S-BH even in the presence of the Bfield so, it is purely due to spin of the BH see Fig. 12.

Hence in this parametric study, we focus on the particle dynamics in the equatorial plan. It will be very important to find if there exists a Carter integral of motion correspond to considered disformal geometry by which we can study the geodesic equations in more general prospective (beyond the case of equatorial orbits). To analyze the possible phenomenons of extracting energy from the BH, possibly utilizing vector interactions need much more study.

Finally, it is quite important to compare the gravitomagnetic effect with MHD driven outflows which is beyond the scope of this paper. The acceleration by Bfield continues at distances much farther from the BH, while the gravitomagnetic effect considered here can only be important near the accretion disk during the initial stages to launch the jet.